# Electronically Tunable Perfect Absorption in Graphene

Seyoon Kim[1,†], Min Seok Jang[1,2,†], Victor W. Brar[1,3,4,†], Kelly W. Mauser[1], and Harry A. Atwater[1,3,*]

* haa@caltech.edu

† Equally contributed authors

1. Thomas J. Watson Laboratory of Applied Physics, California Institute of Technology, Pasadena, CA 91125, United States

2. School of Electrical Engineering, Korea Advanced Institute of Science and Technology, Daejeon 34141, Republic of Korea

3. Kavli Nanoscience Institute, California Institute of Technology, Pasadena, CA 91125, United States

4. Department of Physics, University of Wisconsin-Madison, Madison, WI 53711, United States

**Abstract**

Graphene nanostructures that support surface plasmons have been utilized to create a variety of dynamically tunable light modulators, motivated by theoretical predictions of the potential for unity absorption in resonantly-excited monolayer graphene sheets. Until now, the generally low efficiencies of tunable resonant graphene absorbers have been limited by the mismatch between free-space photons and graphene plasmons. Here, we develop nanophotonic structures that overcome this mismatch and demonstrate electronically tunable perfect absorption achieved with patterned graphenes covering less than 10% of the surface. Experimental measurements reveal 96.9% absorption in the graphene plasmonic nanostructure at 1,389 cm$^{-1}$, with an on/off modulation efficiency of 95.9% in reflection. An analytic effective surface admittance model elucidates the origin of perfect absorption, which is design for critical coupling between free-space modes and the graphene plasmonic nanostructures.

Graphene plasmonic nanostructures have been subjects of intensive research due to the extreme confinement of light in graphene plasmonic resonators. Recently they have also been proposed as candidates for tunable amplitude and phase modulation of THz and mid-infrared light, via a tuning of the carrier density and permittivity under electrostatic gate control. The high confinement factor of graphene plasmons allow for the creation of highly miniaturized [1-6] and active optical elements [7-21]. Tantalizingly, simple theoretical schemes have predicted 100% absorption in resonantly-excited graphene structures [15, 16].

Despite these exciting predictions and attractive features of tunable graphene nanostructures, a major obstacle for graphene plasmonic nanostructures has been the relatively inefficient coupling between free-space photons and graphene plasmons, which has resulted in low device efficiency. This weak interaction between graphene and radiation in free space is due to the inherent thinness, as well as the large wavevector mismatch between graphene plasmons and free-space photons [1, 5, 6]. Another key obstacle has been the low carrier mobility in processed graphene samples as compared with the high mobilities achievable in pristine or passivated and unpatterned graphene sheets [22-26], which have been assumed in theoretical works predicting unity absorption. To circumvent these issues, large chemical doping [7, 10, 13], carefully designed substrates [14-16], and integrated noble metal plasmonic structures [17-21] have been utilized to increase coupling of radiation to graphene plasmonic resonators. In spite of these efforts, not even 50% absorption in graphene has been realized in the mid-infrared.

Here, we develop graphene plasmonic nanostructures that exhibit dramatically higher electronically tunable resonant absorption, made possible by tailored nanostructure designs. First, we utilize lower permittivity substrates to improve resonant absorption in graphene plasmonic ribbon (GPRs) through better wavevector matching between graphene plasmons and free-space photons. Second, we efficiently focus radiation to the GPRs using noble metal plasmonic subwavelength metallic slit arrays, which increase the electric field strength near the GPRs and improve oscillator strength. Third, we design structures in which metal structures situated near GPR dipole resonators give rise to complementary image dipole resonators, further augmenting the light-matter interactions.

Unlike other perfect absorbers solely relying on noble metal plasmonic effects [17, 27], our structures create perfect absorption in the graphene itself by utilizing graphene plasmonic resonances, providing an ideal platform for tunable strong light-matter interactions. Of note, this perfect absorption is achieved with patterned graphenes covering less than 10% of the surface, while unpatterned graphene sheets exhibit low single-pass absorption (~2.3%).

For our graphene plasmonic nanostructures, an effective surface admittance theoretical model serves to guide our designs, and provides insight by revealing that perfect absorption is achieved when the graphene plasmonic nanostructure is carefully tailored to induce critical coupling to free space, i.e., matching of the admittance of graphene resonators to free space. This theoretical model reveals that perfect absorption is no longer limited by low graphene carrier mobility. Finally, reflection measurements for prototype structures demonstrate that electronically tunable resonant absorption can be increased from 24.8% to 96.9 % at 1,389 cm$^{-1}$, with an on/off modulation efficiency of 95.9% in reflection.

Figure 1 shows the three reflection-type light modulator structures based on GPRs. All structures utilize a SiO$_2$/SiN$_x$ membrane with a back reflector as the base substrate, which create a "Salisbury screen" effect to enhance absorption in the GPRs [15, 16]. The type A structure depicted in Fig. 1A consists of periodically arrayed GPRs on the SiO$_2$/SiN$_x$/Au substrate. The type B and C structures in Figs. 1B and C have GPRs coupled to subwavelength metallic slits, with the GPRs located inside the metallic slits. In the type B structure, the GPR is located at the center of the metallic slit, while one side of the GPR is at the center of the slit, and another end is buried beneath the metallic stripe in the type C structure. The total area for each structure is about 75 μm×75 μm, and scanning electron microscope images are shown in Figs. 1D-F for all structures.

The geometry of all structures were chosen such that they would display maximum absorption at 1,356 cm$^{-1}$ with the graphene Fermi level below -0.6 eV (hole doping regime), similar to the conditions for enhanced absorption that we previously reported for a conventional GPR structure on a SiN$_x$/Au substrate [16], which we refer to here as the type 0 structure. In the type A, B, and C structure, SiO$_2$ layers are incorporated to reduce wavevector mismatch between graphene plasmons and free-space photons, which leads to stronger resonances in the GPRs. Varying the substrate and surface environment can yield tangible benefits towards the goal of achieving perfect absorption in electrostatically gated graphene structures with lower graphene carrier mobility.

The Salisbury screen structure in the type 0 and A structures enhances the electric field intensity around GPRs via a Fabry-Perot interference, and thus induces strongly resonant absorption in GPRs [15, 16, 28]. While the back reflector increases the field strength by a factor of 4, larger improvements can be achieved with carefully designed noble metal plasmonic antennas that efficiently focus radiation. As a result, the type B and C slits provide a much stronger field enhancement compared to type A structure, exhibiting enhancement factors of 147 and 226, respectively. In the type B and C structures, the SiN$_x$ thickness is adjusted to 480 nm because the Salisbury screen quarter wavelength resonance condition is no longer only determined by the dielectric material thickness but also by an additional surface inductance from the subwavelength metallic slits, as discussed below.

In addition to enhancing field strength, the metallic slits mimic the interaction between adjacent GPRs to create image GPRs inside the metallic stripes shown in Figs. 1B and C. As a result, the metallic stripes effectively create collective modes, such that the GPRs in all type of structures are equivalent, but the fields are many times stronger in the type B and C structures.

Figure 2A summarizes the calculated absorption in each structure at 1,356 cm$^{-1}$ as a function of graphene hole mobility at the optimal graphene Fermi levels. The graphene hole mobilities required to achieve perfect absorption are 3,174 cm$^2$V$^{-1}$s$^{-1}$, 2,271 cm$^2$V$^{-1}$s$^{-1}$, 613 cm$^2$V$^{-1}$s$^{-1}$, and 315 cm$^2$V$^{-1}$s$^{-1}$ for the type 0, A, B and C structures, respectively. Compared to type 0 and A structures, perfect absorption is achievable at lower graphene hole mobility in the type B and C structures due to the enhanced field around the GPRs. The type C slit configuration exhibits a larger field enhancement than the type B slit because the narrower slit in type C more effectively confines the light. As a result, the type C structure achieves perfect absorption at the lowest graphene hole mobility. Figure 2B illustrates tunable absorption spectra as a function of graphene Fermi level with $\mu_h$=315 cm$^2$V$^{-1}$s$^{-1}$; the maximum absorption values for the type 0, A, B, and C structures are 50.1%, 50.8%, 91.7%, and 100%, respectively.

While full-wave simulations indicate a large absorption enhancement in the proposed structures, it is desirable to develop an analytical model in order to fully understand the underlying photonic concepts, as a guide to design. Since the thickness of the graphene plasmonic nanostructures are much thinner than the free-space wavelength, the top layers can be modeled by an effective surface admittance [16, 17, 28]. The reflection coefficient from the entire structure is derived as

$$r = -\frac{\tilde{Y}_s + \tilde{Y}_{sub} - 1}{\tilde{Y}_s + \tilde{Y}_{sub} + 1} = -\frac{\tilde{Y}_L - 1}{\tilde{Y}_L + 1} \qquad (1)$$

where $\tilde{Y}_{sub}$ is a substrate admittance determined by the substrate geometry. Eq. (1) shows perfect absorption is achieved when the surface admittance $\tilde{Y}_s$ is equal to $1 - \tilde{Y}_{sub}$, and the dotted black line in Fig. 2C corresponds to the $1 - \tilde{Y}_{sub}$, which we term a critical line, as a function of the SiN$_x$ thickness ($d_{SiNx}$).

In the surface admittance chart of Fig. 2C for the type C structure with $\mu_h$=315 cm$^2$V$^{-1}$s$^{-1}$, perfect absorption is achieved at a crossing of the surface admittance line and the critical line, and the maximum surface conductance is strongly dependent on the graphene hole mobility. For a critical graphene hole mobility ($\mu_{h,c}$), the surface admittance and the critical lines form a single critical coupling point at a critical substrate thickness, where perfect absorption occurs. If the graphene hole mobility is lower than the critical graphene hole mobility, the surface admittance line does not cross the critical line, and it corresponds to an under-coupled regime. When $\mu_h > \mu_{h,c}$, two critical coupling points exist and deviate from the critical substrate thickness [16]. In this regime, the resonance at the critical substrate thickness is over-coupled, which explains why the absorption declines after perfect absorption point in Fig. 2A.

Critical coupling can be interpreted as an admittance matching condition. Presenting the load admittance $\tilde{Y}_L$, represented by $\tilde{Y}_s + \tilde{Y}_{sub}$, the critical coupling condition thus corresponds to matching the load admittance to the free-space admittance. Here, the role of the graphene plasmonic nanostructure is to adjust the load admittance so that $\text{Re}(\tilde{Y}_s) = 1$, and the non-zero imaginary part of the $\tilde{Y}_s$ induced by coupling to the subwavelength metallic slits, or the net susceptance, is compensated by the substrate admittance, matching the load admittance to free space.

The surface admittance chart of all structures with $\mu_h$=315 cm$^2$V$^{-1}$s$^{-1}$ is shown in Fig. 2D, and the type C structure exhibits the largest surface conductance. In addition, we observe that the type B and C structures are more inductive (i.e., more negative surface susceptance) than the type 0 and A structures because of the strong noble metal plasmonic resonance induced by the subwavelength metallic slits [29, 30]. Assuming there is no absorption in the dielectric stack and assuming a perfect back reflector, the real part of the substrate admittance becomes zero, and only the imaginary part of the substrate admittance varies, depending on the substrate thickness. For the weak scattering interactions with small $\text{Im}(\tilde{Y}_s)$ exhibited by the type 0 and A structures, the admittance matching condition is satisfied when the thickness of the substrate is equal to a quarter of the wavelength. On the other hand, in the type B and C structures, the $\text{Im}(\tilde{Y}_s)$ is fairly large even when the graphene resonators are off resonance because radiation coupled by the metallic slits significantly advances the phase of light passing through the slits. Due to this abrupt phase advance, admittance matching occurs for thinner substrates compared to weakly scattering Salisbury screens.

Figure 3 illustrates individual absorption in each component at the perfect absorption condition, and shows that most absorption (~95%) occurs in the GPRs. Although the graphene coverage areas in the type B and C structures are less than 10%, their light absorption capacities significantly overwhelm that of an unpatterned graphene sheet. In addition, the coupled noble metal plasmonic structures barely contribute to the total absorption compared with other perfect absorbers [17, 27]. This graphene-dominant absorption is also predictable in the surface admittance chart. In Figs. 2C and D, most of the surface conductance of the graphene nanostructure is achieved by the graphene plasmonic resonance, while the surface conductance is very low at off-resonance conditions, which indicates the low background absorption without the graphene plasmonic resonance.

We can further lower the graphene hole mobility required to achieve perfect absorption by optimizing the geometry of the type C structure. By tailoring the geometry of the type C structure, we achieve perfect absorption with $\mu_h$=200 cm$^2$V$^{-1}$s$^{-1}$. Other designs could allow for perfect absorption with even lower graphene hole mobility, which indicates that perfect absorption in the proposed graphene plasmonic nanostructures is no longer limited by the low graphene carrier mobility.

To demonstrate enhanced electronically-tunable absorption in GPRs coupled to subwavelength metallic slits, we measured the absorption ($A$=1-$R$) for type A structures fabricated on SiO$_2$ 150 nm/SiN$_x$ 1 μm, and type B and C structures fabricated on SiO$_2$ 150 nm/SiN$_x$ 500 nm in a Fourier transform infrared (FTIR) microscope with a polarizer. The modulation efficiencies in reflection are calculated by $\eta_R$=1-$R/R_{\max}$. Here, $R$ corresponds to the gate voltage-dependent reflectance, and $R_{\max}$ is the reflectance when the absorption is minimized at a given graphene Fermi level between the interband absorption and the graphene plasmonic resonance.

**Table 1 | Summary of measurement results**

|                                  | Type A              | Type B              | Type C              |
| -------------------------------- | ------------------- | ------------------- | ------------------- |
| **Frequency**                    | 1,400 cm$^{-1}$     | 1,389 cm$^{-1}$     | 1,407 cm$^{-1}$     |
| **Maximum absorption / $E_F$**   | 52.4% / -0.550 eV   | 96.9% / -0.568 eV   | 94.8% / -0.560 eV   |
| **Minimum absorption / $E_F$**   | 14.0% / -0.316 eV   | 24.8% / -0.262 eV   | 29.6% / -0.262 eV   |
| **Modulation efficiency ($\eta_R$)** | 44.6%           | 95.9%               | 92.6%               |

Figures 4A-C show the gate voltage-dependent tunable absorption spectra, and the corresponding modulation efficiencies are shown in Figs. 4D-F, and Table 1 summarizes the measurement results. In type B structures, we also observed the higher-order graphene plasmonic resonance mode [8, 11], which is not easily observable in the type 0 and A structures for low graphene hole mobility. Figure 4G summarizes the absorption measurements as a function of graphene Fermi level, and corresponding modulation efficiencies are presented in Fig. 4H. The measurement results indicate that the subwavelength metallic slits significantly enhance absorption in the GPRs, and both the type B and C structures display nearly perfect absorption. In addition, the type A structure reported here enables higher tunability than our previously measured type 0 structure [16], which indicates that the low permittivity substrate improves coupling efficiency between free-space photons and graphene plasmons.

To analyze the measurement data, we adjusted the graphene hole mobilities in our simulations to match line shapes of the experimental modulation efficiency spectra, and the simulated maximum modulation efficiencies were fitted to the measured ones [21]. The fitting results reveal that the type A structure is in the under-coupled regime, and the type B structure is close to the critical coupling condition, which results in nearly perfect absorption in the later. The type C structure is expected to be in the over-coupled regime (or, the load admittance exceeds air), which explains the lower absorption in the type C structure than in the type B structure in experimental measurements. In order to enhance the resonant absorption in the type C structure, the graphene hole mobility should be decreased rather than increased to achieve the critical coupling.

Our prototype structures demonstrate that light modulation in graphene is no longer limited by low-quality graphene. These results open the door to highly efficient graphene plasmonic metasurfaces for active infrared optical components, such as modulators, phased arrays, and thermal radiation management. Our design approach using the theoretical surface admittance model also forms the basis for realization of tunable perfect absorption at other frequencies and using other plasmonic materials. Beyond graphene plasmonics, we present state-of-the-art and versatile platforms which can be incorporated with other materials suffering from inherently weak light-matter interactions.

## Acknowledgements

This work was supported by US Department of Energy (DOE) Office of Science grant DE-FG02-07ER46405 (S.K., K.W.M., and H.A.A.), by the Multidisciplinary University Research Initiative Grant, Air Force Office of Scientific Research MURI, Grant No. FA9550-12-1-0488 (V.W.B.), and by the Creative Materials Discovery Program through the National Research Foundation of Korea (NRF) funded by the Ministry of Science, ICT and Future Planning (NRF-2016M3D1A1900038, M.S.J). S.K. acknowledges support by a Samsung Scholarship. The authors thank G. Rossman for assistance with the FTIR microscope and W.-H. Lin for assistance with fabrication.

## Author contributions

S.K. and H.A.A. conceived the ideas. S.K., M.S.J., and K.W.M. performed the simulations and formulated the analytic model. S.K. fabricated the sample. S.K., V.W.B., K.W.M. performed measurements and data analysis. All authors co-wrote the paper, and H.A.A. supervised the project.

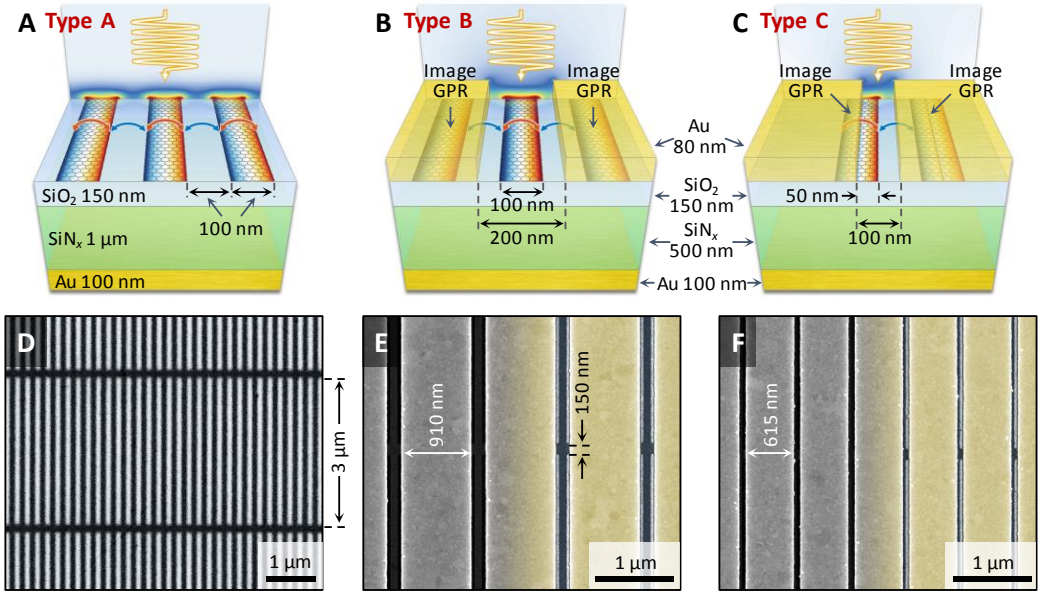

**Figure 1. Structure geometry.** Schematic of (**A-C**) type A, B, and C structures, respectively, and (**D-F**) corresponding scanning electron microscope images (false color). In (A-C), panels at the back side give the out-of-plane electric field distributions, and $E_z$ distributions in graphene are overlapped on graphene plasmonic ribbons (GPRs). The image GPRs in (B) and (C) are created by metallic stripes which operate as mirrors creating virtual GPRs corresponding to the GPRs located in the slits. In the scanning electron microscope images, the dark stripes are GPRs. In panels (E) and (F), the left side of each image is the original SEM image, whereas contrast- and color-adjusted scanning electron microscope images are overlapped on the right side to enhance visibility of the GPRs. To prevent electrical disconnection, GPRs have 150 nm wide bridges, and the length of the GPR strip is 3 μm.

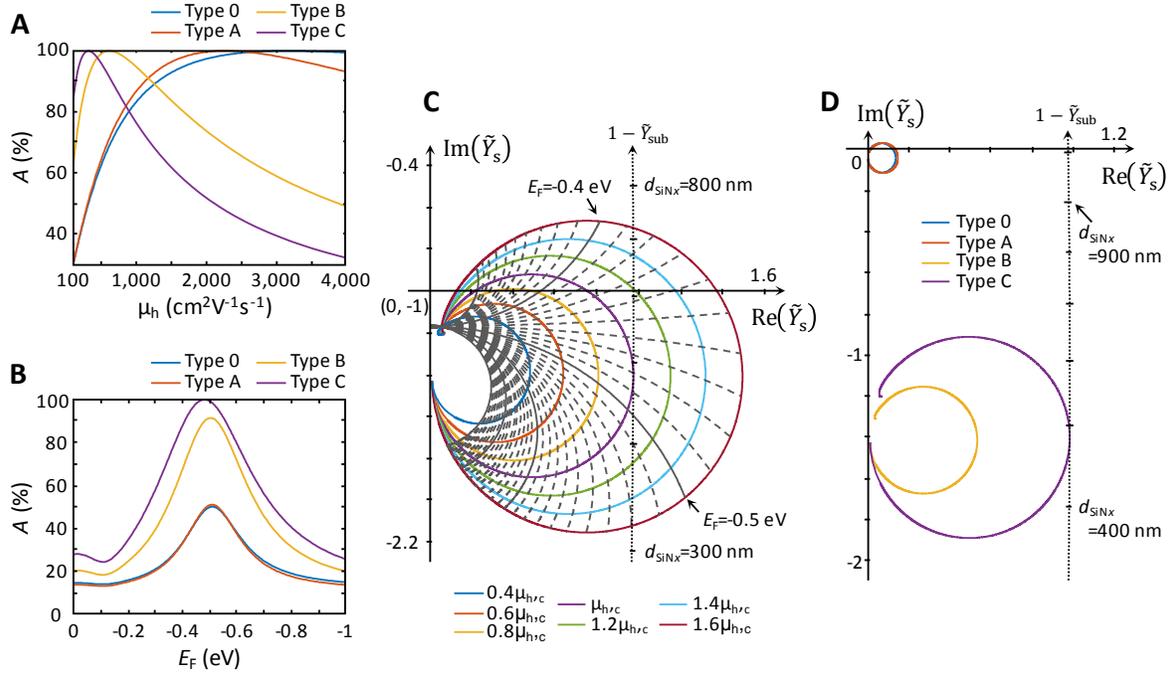

**Figure 2. Theoretical illustration.** (**A**) Absorption in each structure as a function of graphene hole mobility ($\mu_h$). (**B**) Tunable absorption in each structure as a function of graphene Fermi level ($E_F$) for $\mu_h$=315 cm$^2$V$^{-1}$s$^{-1}$. (**C**) Surface admittance chart of the type C structure calculated from full-wave simulations with critical graphene hole mobility ($\mu_{h,c}$) of 315 cm$^2$V$^{-1}$s$^{-1}$; also shown are circles for various graphene hole mobilities ($\mu_h$). The surface admittances are calculated for graphene Fermi levels from 0 eV to -20 eV, and the equi-$E_F$ lines from -0.3 eV to -0.8 eV with 0.01 eV steps (dotted grey lines) and 0.1 eV steps (solid grey lines). (**D**) Surface admittance chart of all structures with $\mu_h$=315 cm$^2$V$^{-1}$s$^{-1}$.

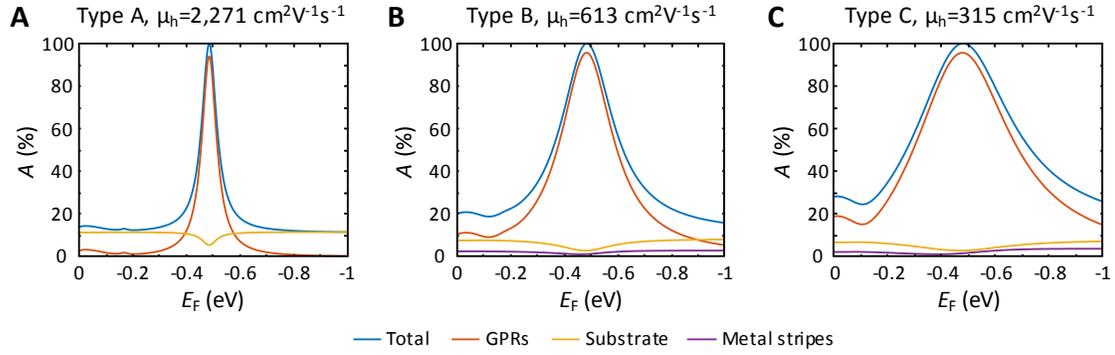

**Figure 3. Absorption in each component.** Calculated absorption in each component at the perfect absorption (**A**) in the type A structure, (**B**) in the type B structure, and (**C**) in the type C structure. The substrate absorption incudes the absorptions in SiO$_2$, SiN$_x$, and Au reflector. The maximum absorptions in the graphene plasmonic ribbons (GPRs) are 94.5%, 96.1%, and 95.9% in the type A, B, and C structures, respectively.

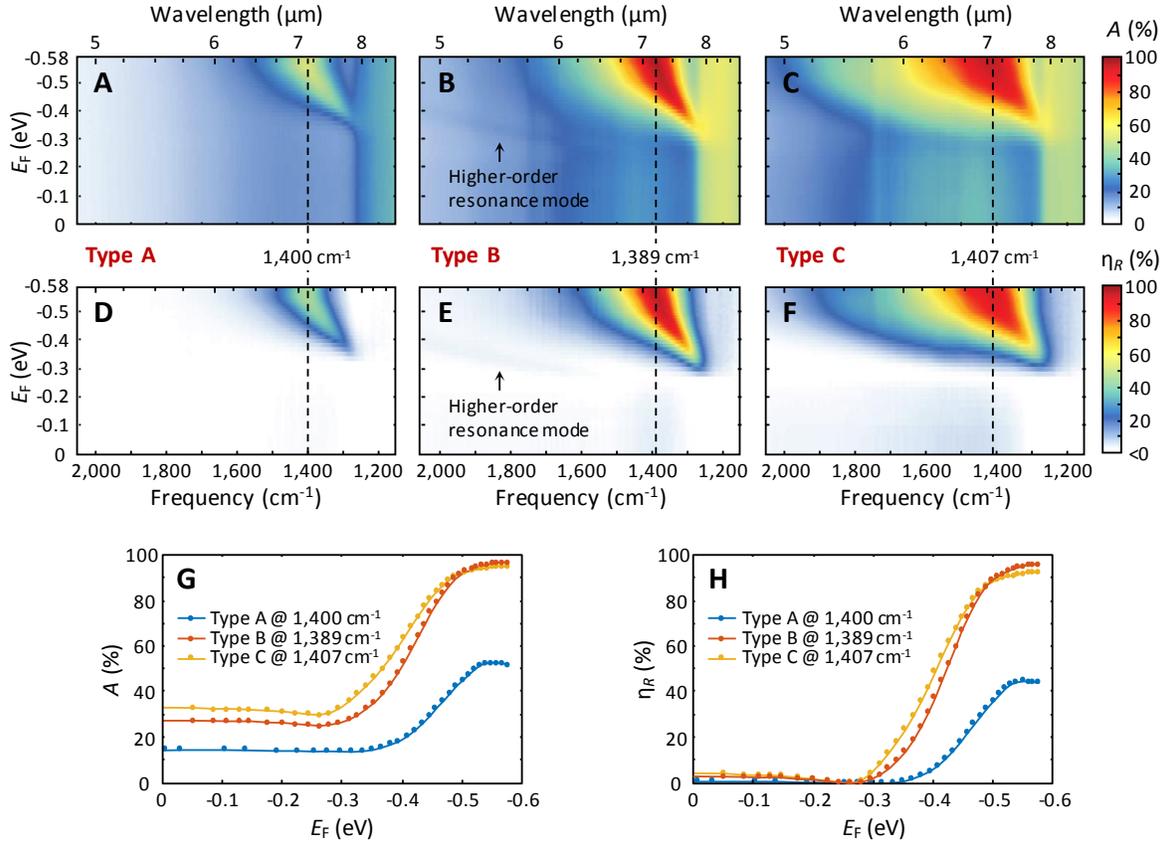

**Figure 4. Experimental results.** (**A**-**C**) Gate voltage-dependent tunable resonant absorption spectra in the type A, B, and C structures, respectively, and (**D**-**F**) corresponding modulation efficiencies ($\eta_R$). (**G**) Absorption and (**H**) modulation efficiency as a function of graphene Fermi level ($E_F$) at the frequency for maximum absorption in each structure.

13